\newif\iflongversion
\longversiontrue %

\documentclass{ifacconf}

\usepackage{graphicx}      %
\usepackage{natbib}        %

\usepackage{url}

\iflongversion
\makeatletter
\let\old@ssect\@ssect %
\makeatother

\usepackage{hyperref}
\usepackage{doi}

\makeatletter
\def\@ssect#1#2#3#4#5#6{%
  \NR@gettitle{#6}%
  \old@ssect{#1}{#2}{#3}{#4}{#5}{#6}%
}
\makeatother
\fi

\usepackage[percent]{overpic}

\usepackage{subcaption}
\usepackage{enumerate}

\usepackage{bm}
\usepackage[bbsets,Dfprime,setrelation]{math}

\usepackage{wasysym}
\usepackage[prefixflatinterpret,bracketmodalinterpret,fixformat,silentconst,sidenotecalculus,longseqcontext,seqinsist]{logic}
\usepackage[prefixflatinterpret,bracketmodalinterpret,fixformat,silentconst,differentialdL,simplenames]{dL}
\def\leftrule{L}%
\def\rightrule{R}%
\newcommand{\bebecomes}{\mathrel{::=}}
\newcommand{\alternative}{~|~}

\usepackage{xcolor}
\usepackage{paralist}

\newcommand{\progA}{\alpha}
\newcommand{\progB}{\beta}
\newcommand{\etermA}{e}
\newcommand{\fvarA}{\phi}
\newcommand{\fvarB}{\psi}
\newcommand{\rfvar}{P}
\newcommand{\rifvar}{I}

\newcommand{\rrfvar}{R}

\newcommand{\arbswitch}{\alpha_{\texttt{arb}}}
\newcommand{\stateswitch}{\alpha_{\texttt{state}}}
\newcommand{\slowswitch}{\alpha_{\texttt{slow}}}
\newcommand{\ctrlswitch}{\alpha_{\texttt{ctrl}}}
\newcommand{\fastswitch}{\alpha_{\texttt{fast}}}

\newcommand{\ifthen}[2]{\texttt{if}(#1)\{#2\}}

\newcommand{\sigfam}{\mathcal{P}}
\newcommand{\lterm}{v}

\newcommand{\I}{\dLint[state=\omega]}
\newcommand{\It}{\dLint[state=\nu]}

\newsavebox{\Rval}%
\sbox{\Rval}{$\scriptstyle\mathbb{R}$}
\newsavebox{\Rvalext}%
\sbox{\Rvalext}{$\scriptstyle\mathbb{R}_{\exp,\sin,\cos}$}
\newsavebox{\Rvalexp}%
\sbox{\Rvalexp}{$\scriptstyle\mathbb{R}_{\exp}$}
\relax

  \newdimen\linferenceRulehskipamount%
  \newdimen\lcalculuscollectionvskipamount%
\definecolor{vblue}{rgb}{.1,.15,.62}
\definecolor{vgray}{rgb}{.35,.35,.35}
\renewcommand*{\irrulename}[1]{\text{\textcolor{vgray}{\upshape#1}}}%
\renewcommand{\linferenceRuleNameSeparation}{\hspace{5pt}}
\renewcommand{\linferenceRuleSidenoteSeparation}{\quad}
\renewcommand{\linferenceRuleRefSeparation}{, }

\usepackage{tikz}
\usetikzlibrary{arrows}

\newcommand*\circled[1]{\tikz[baseline=(char.base)]{
            \node[shape=circle,draw,inner sep=1pt] (char) {#1};}}

\renewcommand*{\lie}[3][]
{\mathcal{L}_{#2}^{\ifthenelse{\equal{#1}{}}{}{^{\left(#1\right)}}}(#3)}
\renewcommand*{\lied}[3][]{\overset{\bm .}{#3}\ifthenelse{\equal{#1}{}}{}{{}^{(#1)}}}

\newcommand{\siglie}[3][]{\overset{\bm .}{#3}{}^{\Dostar{#1}}_{#2}}

\renewcommand{\Dostar}[1]{\ifthenelse{\equal{#1}{}}{(*)}{-(*)}}

\newcommand{\sigliesai}[3][]{\siglie[#1]{#2}{#3}}

\newcommand{\solvar}{\varphi}
\newcommand{\solvarr}{\psi}

\usepackage{prettyref}
\newcommand{\rref}[2][]{\prettyref{#2}}
\newrefformat{sec}{Section\,\ref{#1}}
\newrefformat{subsec}{Section\,\ref{#1}}
\newrefformat{def}{Def.\,\ref{#1}}
\newrefformat{eqn}{Equation\,\ref{#1}}
\newrefformat{thm}{Theorem\,\ref{#1}}
\newrefformat{fact}{Fact\,\ref{#1}}
\newrefformat{prop}{Proposition\,\ref{#1}}
\newrefformat{lem}{Lemma\,\ref{#1}}
\newrefformat{cor}{Corollary\,\ref{#1}}
\newrefformat{ex}{Example\,\ref{#1}}
\newrefformat{tab}{Table\,\ref{#1}}
\newrefformat{fig}{Fig.\,\ref{#1}}
\newrefformat{subfig}{Fig.\,\ref{#1}}
\newrefformat{case}{case\,\ref{#1}}
\newrefformat{foot}{Footnote\,\ref{#1}}
\newrefformat{itm}{\ref{#1}}

\newrefformat{app}{Appendix\,\ref{#1}}

\begin{document}
\begin{frontmatter}
\iflongversion
\endNoHyper %
\fi

\title{Switched Systems as Hybrid Programs\thanksref{footnoteinfo}}

\thanks[footnoteinfo]{
This research was sponsored by the AFOSR under grant number FA9550-16-1-0288.
The first author was also supported by A*STAR, Singapore.
\iflongversion
The views and conclusions contained in this document are those of the authors and should not be interpreted as representing the official policies, either expressed or implied, of any sponsoring institution, the U.S. government or any other entity.\\
\copyright~2021 the authors. This work has been accepted to IFAC for publication under a Creative Commons Licence CC-BY-NC-ND.
\fi
}

\author{Yong Kiam Tan}\qquad\author{Andr\'e Platzer}

\address{Computer Science Department, Carnegie Mellon University,
  Pittsburgh, USA (e-mail: \{yongkiat,aplatzer\}@cs.cmu.edu)}

\begin{abstract}                %
Real world systems of interest often feature interactions between discrete and continuous dynamics.
Various hybrid system formalisms have been used to model and analyze this combination of dynamics, ranging from mathematical descriptions, e.g., using impulsive differential equations and switching, to automata-theoretic and language-based approaches.
This paper bridges two such formalisms by showing how various classes of switched systems can be modeled using the language of hybrid programs from differential dynamic logic (\dL).
The resulting models enable the formal specification and verification of switched systems using \dL and its existing deductive verification tools such as \KeYmaeraX.
Switched systems also provide a natural avenue for the generalization of \dL's deductive proof theory for differential equations.
The completeness results for switched system invariants proved in this paper enable effective safety verification of those systems in \dL.

\iflongversion
~
\fi
\end{abstract}

\iflongversion
\else
\fi

\end{frontmatter}

\section{Introduction}
\label{sec:introduction}

The study of \emph{hybrid systems}, i.e., mathematical models that combine discrete and continuous dynamics, is motivated by the need to understand the hybrid dynamics present in many real world systems~\citep{DBLP:books/sp/Liberzon03,DBLP:books/sp/Platzer18}.
Various formalisms can be used to describe hybrid systems, for example,
impulsive differential equations~\citep{10.2307/j.ctt7zvqf0};
switched systems~\citep{DBLP:books/sp/Liberzon03,SunGe};
hybrid time combinations of discrete and continuous dynamics~\citep{4806347,10.2307/j.ctt7s02z};
hybrid automata~\citep{DBLP:conf/lics/Henzinger96};
and language-based models~\citep{DBLP:journals/tcs/RonkkoRS03,DBLP:conf/aplas/LiuLQZZZZ10,DBLP:books/daglib/0025392,DBLP:books/sp/Platzer18}.
These formalisms differ in their generality and in how the discrete-continuous dynamical combination is modeled, e.g., ranging from differential equations with discontinuous right-hand sides, to combinators that piece together discrete and continuous programs.
Consequently, different formalisms may be better suited for different hybrid system applications and it is worthwhile to explore connections between different formalisms in order to exploit their various strengths for a given application.

A \emph{switched system} consists of a family of continuous ordinary differential equations (ODEs) together with a discrete switching signal that prescribes the active ODE the system follows at each time.
These models are commonly found in control designs where appropriately designed switching can be used to achieve control goals that cannot be achieved by purely continuous means~\citep{DBLP:books/sp/Liberzon03}.

Differential dynamic logic (\dL)~\citep{DBLP:books/daglib/0025392,DBLP:books/sp/Platzer18} provides the language of \emph{hybrid programs}, whose hybrid dynamics arise from combining discrete programming constructs with continuous ODEs.
This combination yields a rich and flexible language for describing hybrid systems, e.g., with event- or time-triggered design paradigms.

This paper shows how various classes of switched systems can be fruitfully modeled in the language of hybrid programs.
The contributions are as follows:
\begin{enumerate}
\item Important classes of switched systems are modeled as hybrid programs in Sections~\ref{sec:arbstateswitching}--\ref{sec:timectrlswitching}.
Subtleties associated with those models are investigated, along with methods for detecting and avoiding those pitfalls.
\item Completeness results for differential equation invariants in \dL~\citep{DBLP:journals/jacm/PlatzerT20} are extended to invariants of switched systems, yielding an effective technique for proving switched system safety.
\end{enumerate}

These contributions enable sound deductive verification of switched systems in \dL and they lay the groundwork for further development of proof automation for switched systems, such as in the \KeYmaeraX~\citep{DBLP:conf/cade/FultonMQVP15} hybrid systems prover based on \dL.
To demonstrate the versatility of the proposed hybrid program models,~\rref{sec:stability} uses \KeYmaeraX to formally verify stability for several switched system examples using standard Lyapunov function techniques~\citep{DBLP:books/sp/Liberzon03}.
\iflongversion
All proofs are available in~\rref{app:proofs}.
\else
All proofs are available in the supplement~\citep{DBLP:journals/corr/abs-2101-06195}.
\fi

\section{Background}
\label{sec:background}

This section informally recalls differential dynamic logic (\dL) and the language of hybrid programs used to model switched systems in Sections~\ref{sec:arbstateswitching} and~\ref{sec:timectrlswitching}.
Formal presentations of \dL are available elsewhere~\citep{DBLP:books/daglib/0025392,DBLP:journals/jar/Platzer17,DBLP:books/sp/Platzer18}.

\subsection{Hybrid Programs}

The language of \emph{hybrid programs} is generated by the following grammar, where $x$ is a variable, $\etermA$ is a \dL term, e.g., a polynomial over $x$, and $\ivr$ is a \dL formula.
\[
	\progA,\progB~\bebecomes~\pumod{x}{e} \alternative \ptest{\ivr} \alternative \pevolvein{\D{x}=\genDE{x}}{\ivr} \alternative \progA ; \progB \alternative \pchoice{\progA}{\progB} \alternative \prepeat{\progA}
\]

\begin{figure}[h]
\centering

\begin{overpic}[width=\columnwidth,tics=10,clip,trim=500 320 500 280]{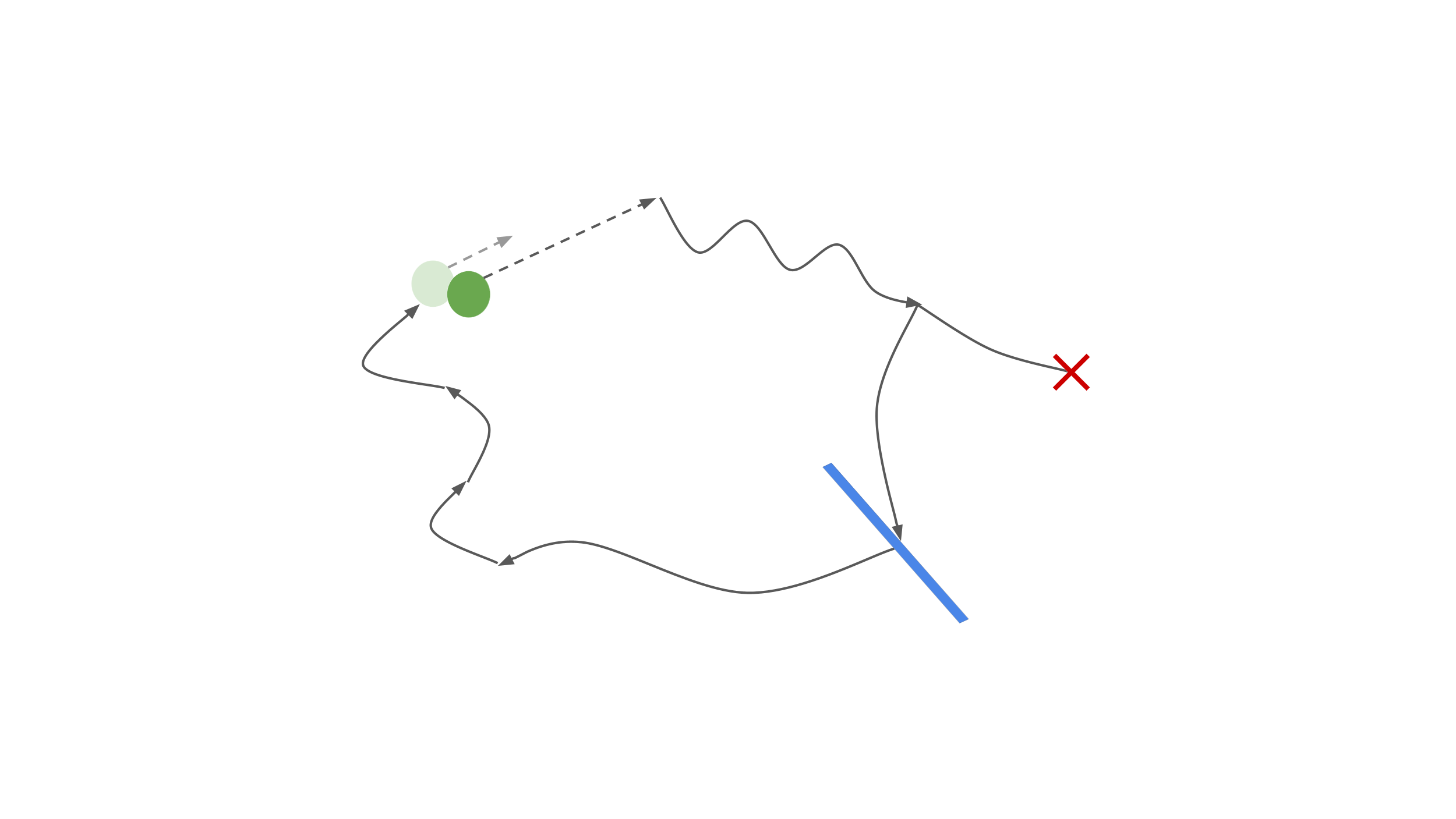}
 \put (28,51) {\small A) $\pumod{x}{e};\D{x}=\genDE{x}$}
 \put (67,40) {\small B) $\pchoice{\D{x}{=}\genDE{x}}{\D{x}{=}g(x)}$}
 \put (45,24) {\small C.i) $\ptest{\ivr}$ (true)}
 \put (76,24) {\small C.ii) $\ptest{\ivr}$ (false)}
 \put (70,13) {\small D) State-dependent}
 \put (75,9) {\small switching}
 \put (59,9) {\small $t=0$}
 \put (50,2) {\small $t=1$}
 \put (40,9) {\small $t=2$}
 \put (24,5) {\small $t\geq\tau$}
 \put (21,13) {\small E) Time-dependent switching}
 \put (0,24) {\small F) Controlled}
 \put (5,20) {\small switching}
 \put (10,44) {\small G) \prepeat{\alpha}}
\end{overpic}

\captionsetup{singlelinecheck=off}
\caption[foo]{The green initial state evolving according to a hybrid program featuring (clockwise from top):
\begin{enumerate}[A]
\item a discrete assignment (dashed line) followed sequentially by continuous ODE evolution (solid line),
\item a choice between two ODEs (\rref{sec:arbswitching}),
\item a test that aborts (red $\times$) system evolutions leaving $\ivr$,
\item switching when the system state crosses the thick blue switching surface (\rref{sec:stateswitching}),
\item switching after time $t \geq \tau$ has elapsed (\rref{sec:timeswitching}),
\item switching control that is designed to drive the system state close to its initial position (\rref{sec:ctrlswitching}), and
\item a loop that repeats system evolution (in lighter colors).
\end{enumerate}}
\label{fig:switchedsys}
\end{figure}

Discrete assignment $\pumod{x}{e}$ sets the value of variable $x$ to that of term $e$ in the current state.
Test $\ptest{\ivr}$ checks that formula $\ivr$ is true in the current state and aborts the run otherwise.
The continuous program $\pevolvein{\D{x}=\genDE{x}}{\ivr}$ continuously evolves the system state by following the ODE $\D{x}=\genDE{x}$ for a nondeterministically chosen duration $t \geq 0$, as long as the system remains in the domain constraint $\ivr$ for all times $0 \leq \tau \leq t$.
The sequence program $\progA ; \progB$ runs program $\progB$ after $\progA$, the choice program $\pchoice{\progA}{\progB}$ nondeterministically chooses to run either $\progA$ or $\progB$, and the loop program $\prepeat{\progA}$ repeats $\progA$ for $n \in \naturals$ iterations where $n$ is chosen nondeterministically.
The nondeterminism inherent in hybrid programs is useful for abstractly modeling real world behaviors~\citep{DBLP:books/sp/Platzer18}.
The evolution of various hybrid programs is illustrated in parts A--C and G of~\rref{fig:switchedsys}.

Notationally, $x = (x_1,\dots,x_n)$ are the state variables of an $n$-dimensional system, so $\pevolvein{\D{x}=\genDE{x}}{\ivr}$ is an autonomous $n$-dimensional system of ordinary differential equations over $x$; the ODE is written as $\D{x}=\genDE{x}$ when there is no domain constraint, i.e., $\ivr \mnodefequiv \ltrue$.
For simplicity, all ODEs have polynomial right-hand sides, \dL terms $e$ are polynomial over $x$, and $\rfvar,\ivr$ are formulas of first-order real arithmetic over $x$; extensions of the term language to Noetherian functions are described in~\cite{DBLP:journals/jacm/PlatzerT20}.
The single-sided conditional \texttt{if} is defined as $\ifthen{\rfvar}{\alpha} \mnodefequiv \pchoice{(\ptest{\rfvar};\alpha)}{(\ptest{\lnot{\rfvar}})}$.
Nondeterministic choice over a finite family of hybrid programs $\alpha_p$ for $p \in \sigfam$, $\sigfam\mnodefequiv \{1,\dots,m\}$ is denoted $\bigcup_{p \in \sigfam}{\alpha_p} \mnodefequiv \pchoice{\alpha_1}{\pchoice{\alpha_2}{\pchoice{\dots}{\alpha_m}}}$.

The formula language of \dL extends first-order logic formulas with dynamic modalities for specifying properties of a hybrid program $\alpha$~\citep{DBLP:journals/jar/Platzer17,DBLP:books/sp/Platzer18}.
The box modality formula $\dbox{\alpha}{\rfvar}$ says that formula $\rfvar$ is true for \emph{all} states reachable by following the nondeterministic evolutions of hybrid program $\alpha$, while the diamond modality formula $\ddiamond{\alpha}{\rfvar}$ says that formula $\rfvar$ is true for \emph{some} reachable state of $\alpha$.
This paper focuses on using box modality formulas for specifying safety properties of hybrid programs.
For example, formula $\rrfvar \limply \dbox{\prepeat{\alpha}}{\rfvar}$ says that initial states satisfying precondition $\rrfvar$ remain in the safe region $\rfvar$ after any number of runs of the loop $\prepeat{\alpha}$.
A key technique for proving safety properties of such a loop is to identify an \emph{invariant} $\rifvar$ of $\alpha$ such that formula $\rifvar \limply \dbox{\alpha}{\rifvar}$ is \emph{valid}, i.e., true in all states~\citep{DBLP:books/sp/Platzer18}.
To enable effective proofs of safety, invariance, and various other properties of interest, \dL provides compositional reasoning principles for hybrid programs~\citep{DBLP:journals/jar/Platzer17,DBLP:books/sp/Platzer18} and a complete axiomatization for ODE invariants~\citep{DBLP:journals/jacm/PlatzerT20}.

\subsection{Switched Systems}
\label{sec:switchedsystems}

A \emph{switched system} is described by the following data:
\begin{enumerate}
\item an open, connected set $D \subseteq \reals^n$ which is the \emph{state space} of interest for the system,
\item a finite (non-empty) family $\sigfam$ of ODEs $\D{x}=f_p{(x)}$ for $p \in \sigfam$, and,
\item for each initial state $\iget[state]{\I} \in D$, a set of \emph{switching signals} $\sigma : [0,\infty) \to \sigfam$ prescribing the ODE $\D{x}=f_{\sigma(t)}{(x)}$ to follow at time $t$ for the system's evolution from $\iget[state]{\I}$.\footnote{
\iflongversion
A more precise definition is given in~\rref{app:proofs}, where the switching signals $\sigma$ are also required to be well-defined~\citep{DBLP:books/sp/Liberzon03,SunGe} so that they model physically realizable switching.
\else%
A more precise definition is given in the supplement~\citep{DBLP:journals/corr/abs-2101-06195}, where the switching signals $\sigma$ are also required to be well-defined~\citep{DBLP:books/sp/Liberzon03,SunGe} so that they model physically realizable switching.
\fi}
\end{enumerate}

Switching phenomena can either be described explicitly as a function of time, or implicitly, e.g., as a state predicate, depending on the real world switching mechanism being modeled.
Several standard classes of switching mechanisms are studied in Sections~\ref{sec:arbstateswitching} and \ref{sec:timectrlswitching}, following the nomenclature from~\cite{DBLP:books/sp/Liberzon03}.
These switching mechanisms are illustrated in parts D--F of~\rref{fig:switchedsys}.

For simplicity, this paper assumes that the state space is $D \mnodefeq \reals^n$.
More general definitions of switched systems are possible but are left out of scope, see~\cite{DBLP:books/sp/Liberzon03}.
For example, $\sigfam$ can more generally be an (uncountably) infinite family and some switched systems may have \emph{impulse effects} where the system state is allowed to make instantaneous, discontinuous jumps during the system's evolution, such as the dashed jump in part A of~\rref{fig:switchedsys}.

\section{Arbitrary and State-Dependent Switching}
\label{sec:arbstateswitching}

\subsection{Arbitrary Switching}
\label{sec:arbswitching}

Real world systems can exhibit switching mechanisms that are uncontrolled, \emph{a priori} unknown, or too complicated to describe succinctly in a model.
For example, a driving vehicle may encounter several different road conditions depending on the time of day, weather, and other unpredictable factors---given the multitude of combinations to consider, it is desirable to have a single model that exhibits and switches between all of those road conditions.
\emph{Arbitrary switching} is a useful paradigm for such systems because it considers \emph{all} possible switching signals and their corresponding system evolutions.
The arbitrary switching mechanism is modeled by the following hybrid program and illustrated in~\rref{fig:arbswitching}.
\begin{equation*}
\arbswitch \mnodefequiv \prepeat{\Big( \bigcup_{p \in \sigfam}{\D{x}=f_p{(x)}} \Big)}
\end{equation*}

\begin{figure}
\centering
\includegraphics[width=0.85\columnwidth]{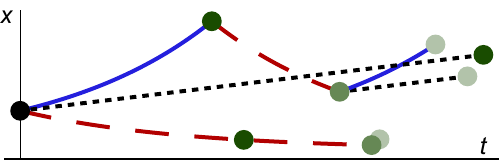}
\caption{Evolution of $\arbswitch$ for $\D{x}=x$ (solid blue), $\D{x}=1$ (dotted black), and $\D{x}=-x$ (dashed red) from the initial state (black circle). Switching steps are marked by green circles and faded colors illustrate progression in loop iterations for the loop operator in $\arbswitch$.}
\label{fig:arbswitching}
\end{figure}

Observe that
\begin{inparaenum}[\it i)]
\item the system nondeterministically chooses which ODE to follow at each loop iteration;
\item it follows the chosen ODE for a nondeterministic duration;
\item each loop iteration corresponds to a switching step and the loop repeats for a finite, nondeterministically chosen number of iterations.
\end{inparaenum}
Two subtle behaviors are illustrated by the bottom trajectory in~\rref{fig:arbswitching}: $\arbswitch$ can switch to the same ODE across a loop iteration or it can \emph{chatter} by making several discrete switches without continuously evolving its state between those switches~\citep{DBLP:journals/tecs/SogokonGJ17}.
These behaviors are harmless for safety verification because they do not change the set of reachable states of the switched system.
Formally, the adequacy of $\arbswitch$ as a model of arbitrary switching is shown in the following proposition.

\begin{prop}
A state is reachable by hybrid program $\arbswitch$ iff it is reachable in finite time by a switched system $\D{x}=f_p(x)$ for $p \in \sigfam$ following a switching signal $\sigma$.
\label{prop:adequacyarbswitching}
\end{prop}

By~\rref{prop:adequacyarbswitching}, the \dL formula $\dbox{\arbswitch}{\rfvar}$ specifies safety for arbitrary switching, i.e., for any switching signal $\sigma$, the system states reached at all times by switching according to $\sigma$ satisfy the safety postcondition $\rfvar$.

\subsection{State-Dependent Switching}
\label{sec:stateswitching}

Arbitrary switching can be constrained by enabling switching to the ODE $\D{x}=f_p{(x)}$ only when the system state belongs to a corresponding domain specified by formula $\ivr_p$.
This yields the \emph{state-dependent switching} paradigm, which is useful for modeling real systems that are either known or designed to have particular switching surfaces.
For the finite family of ODEs with domains $\pevolvein{\D{x}=f_p{(x)}}{\ivr_p}$, $p \in \sigfam$, state-dependent switching is modeled as follows:
\begin{equation*}
\stateswitch \mnodefequiv \prepeat{\Big( \bigcup_{p \in \sigfam}{ \pevolvein{\D{x}=f_p{(x)}}{\ivr_p}} \Big)}
\end{equation*}

Operationally, if the system is currently evolving in domain $\ivr_i$ and is about to leave the domain, it must switch to another ODE with domain $\ivr_j$ that is true in the current state to continue its evolution.
Arbitrary switching $\arbswitch$ is the special case of $\stateswitch$ with no domain restrictions.
The following result generalizes~\rref{prop:adequacyarbswitching} to consider only states reached while obeying the specified domains.

\begin{prop}
A state is reachable by hybrid program $\stateswitch$ iff it is reachable in finite time by a switched system $\D{x}=f_p(x)$ for $p \in \sigfam$ following a switching signal $\sigma$ while obeying the domains $\ivr_p$.
\label{prop:adequacystateswitching}
\end{prop}

The next two results are syntactically provable in \dL and they provide sound and complete invariance reasoning principles for state-dependent (and arbitrary) switching.
Formula $\phi$ is \emph{computable} from a set of inputs iff there is an algorithm that outputs $\phi$ when given those inputs.

\begin{lem}%
Formula $\rifvar$ is an invariant for $\stateswitch$ iff $\rifvar$ is invariant for all constituent ODEs $\pevolvein{\D{x}=f_p{(x)}}{\ivr_p}$, $p \in \sigfam$.
\label{lem:compinvariance}
\end{lem}

\begin{thm}
From input ODEs $\pevolvein{\D{x}=f_p{(x)}}{\ivr_p}$, $p \in \sigfam$ and formula $\rifvar$, there is a computable formula of real arithmetic $\phi$ such that formula $\rifvar$ is invariant for $\stateswitch$ iff $\phi$ is valid.
In particular, invariance for $\stateswitch$ is decidable.
\label{thm:decidability}
\end{thm}

\rref{lem:compinvariance} shows that when searching for an invariant of $\stateswitch$, it suffices to search for a \emph{common} invariant of every constituent ODE.
\rref{thm:decidability} enables sound and complete invariance proofs for systems with state-dependent switching in \dL, relying on \dL's complete axiomatization for ODE invariance and decidability of first-order real arithmetic over polynomial terms~\citep{Tarski}.
These results also extend to Noetherian functions, e.g., exponentials and trigonometric functions, at the cost of losing decidability of the resulting arithmetic~\citep{DBLP:journals/jacm/PlatzerT20}.

\subsection{Modeling Subtleties}
\label{sec:modelsubtleties}

The model $\stateswitch$ as defined above makes no \emph{a priori} assumptions about how the ODEs and their domains $\pevolvein{\D{x}=f_p(x)}{\ivr_p}$ are designed, so results like~\rref{thm:decidability} apply generally to all state-dependent switching designs.
However, state-dependent switching can exhibit some well-known subtleties~\citep{DBLP:books/sp/Liberzon03,DBLP:journals/tecs/SogokonGJ17} and it becomes the onus of modelers to appropriately account for these subtleties.
This section examines various subtleties that can arise in $\stateswitch$ and prescribes sufficient arithmetical criteria for avoiding them;
like~\rref{thm:decidability}, these arithmetical criteria are decidable for systems with polynomial terms~\citep{Tarski}.
As a running example, let the line $x_1=x_2$ be a \emph{switching surface}, i.e., the example systems described below are intended to exhibit switching when their system state reaches this line.

\subsubsection{Well-defined switching.}
First, observe that the domains $\ivr_p$ must cover the entire state space; otherwise, there would be system states of interest where no continuous dynamics is active.
This can be formally guaranteed by deciding validity of the formula \circled{1}: $\lorfold_{p \in \sigfam}{\ivr_p}$.
Next, consider the following ODEs:

\begin{minipage}{0.23\textwidth}
\begin{align*}
&\underbrace{\pevolvein{\D{x_1}=0, \D{x_2} = 1}{x_1 \geq x_2}}_{\pevolvein{\D{x}=f_A(x)}{\ivr_A}} \\
\\
&\underbrace{\pevolvein{\D{x_1}=-1, \D{x_2} = 0}{x_1 < x_2}}_{\pevolvein{\D{x}=f_B(x)}{\ivr_B}} \\
\end{align*}
\end{minipage}
\hfill\begin{minipage}{0.22\textwidth}
\includegraphics[width=0.85\columnwidth,clip,trim=5 5 15 10]{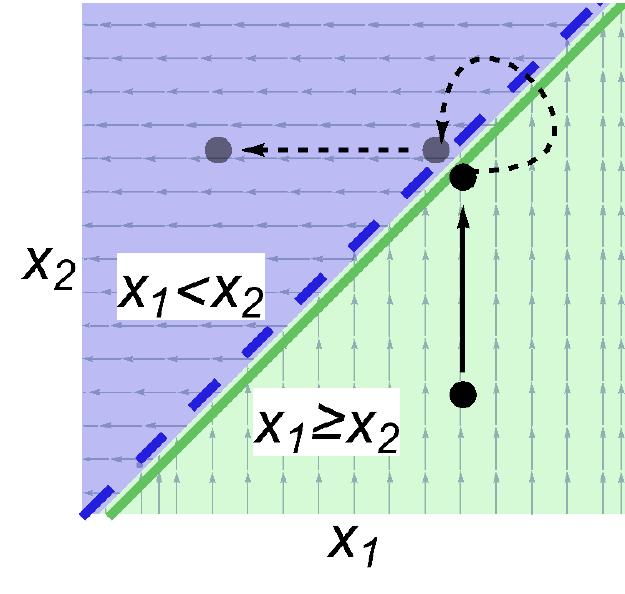}
\end{minipage}

Consider the system evolution starting in $\ivr_A \mnodefequiv x_1 \geq x_2$ illustrated above on the right.
When the system reaches $x_1=x_2$ (the illustration is offset for clarity), it is about to \emph{locally progress} into $\ivr_B \mnodefequiv x_1 < x_2$ by switching to ODE $\D{x}=f_B(x)$ but it gets stuck because it cannot make the infinitesimal jump from $\ivr_A$ to enter $\ivr_B$; augmenting domain $\ivr_B$ to $x_1 \leq x_2$ enables the switch.
More generally, to avoid the need for infinitesimal jumps, domains $\ivr_p$ should be augmented to include states that locally progress into $\ivr_p$ under the ODE $\D{x}=f_p(x)$ and, symmetrically, states that locally exit $\ivr_p$~\citep{DBLP:journals/tecs/SogokonGJ17}.
Local progress (and exit) for ODEs is characterized as follows.

\begin{thm}[\cite{DBLP:journals/jacm/PlatzerT20}]
From input ODE $\pevolvein{\D{x}=f{(x)}}{\ivr}$, there are computable formulas of real arithmetic $\sigliesai{f}{(\ivr)}$, $\sigliesai{-f}{(\ivr)}$ that respectively characterize the states from which $\D{x}=f(x)$ locally progresses into $\ivr$ and those from which it locally exits $\ivr$.
\label{thm:localprogress}
\end{thm}

By~\rref{thm:localprogress}, to avoid the stuck states exemplified above for ODEs $\pevolvein{\D{x}=f_p(x)}{\ivr_p}$, $p \in \sigfam$ in $\stateswitch$, it suffices to decide validity of the formula \circled{2}: $\sigliesai{f_p}{(\ivr_p)} \lor \sigliesai{-f_p}{(\ivr_p)} \limply \ivr_p$ for each $p \in \sigfam$.
Condition~\circled{2} is syntactically significantly simpler but equivalent to the domain augmentation presented in~\cite{DBLP:journals/tecs/SogokonGJ17} for piecewise continuous models, a form of state-dependent switching.

\subsubsection{Sliding modes.}
The preceding subtlety arose from incomplete domain constraint specifications.
Another subtlety that can arise because of incomplete specification of ODE dynamics is exemplified by the following ODEs:

\begin{minipage}{0.23\textwidth}
\begin{align*}
&\underbrace{\pevolvein{\D{x_1}=0, \D{x_2} = 1}{x_1 \geq x_2}}_{\pevolvein{\D{x}=f_A(x)}{\ivr_A}} \\
\\
&\underbrace{\pevolvein{\D{x_1}=1, \D{x_2} = 0}{x_1 \leq x_2}}_{\pevolvein{\D{x}=f_B(x)}{\ivr_B}} \\
\end{align*}
\end{minipage}
\hfill\begin{minipage}{0.22\textwidth}
\includegraphics[width=0.85\columnwidth,clip,trim=5 5 15 10]{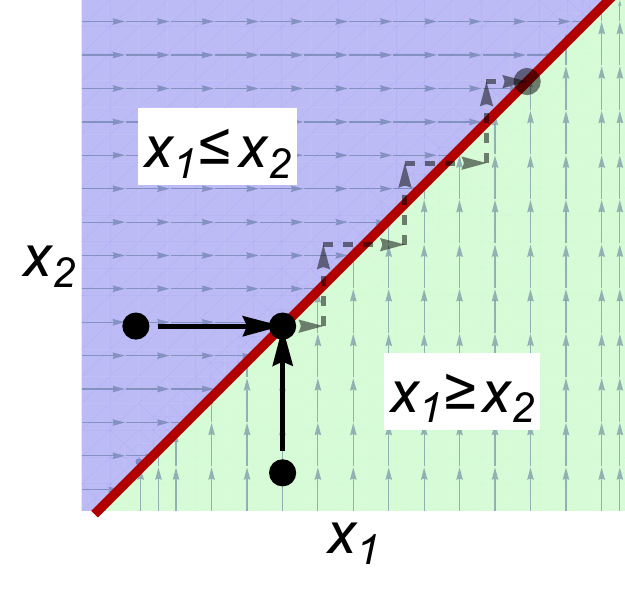}
\end{minipage}

Systems starting in $\ivr_A \mnodefequiv x_1 \geq x_2$ or $\ivr_B \mnodefequiv x_1 \leq x_2$ eventually reach the line $x_1=x_2$ but they then get stuck because the ODEs on either side of $x_1=x_2$ drive system evolution onto the line.
Mathematically, the system enters a \emph{sliding mode}~\citep{DBLP:books/sp/Liberzon03} along $x_1=x_2$; as illustrated above, this can be thought of as infinitely fast switching between the ODEs that results in a new sliding dynamics \emph{along} the switching surface $x_1=x_2$.

When the sliding dynamics can be calculated exactly, it suffices to add those dynamics to the switched system, e.g., adding the sliding dynamics $\pevolvein{\D{x_1}=\frac{1}{2}, \D{x_2} = \frac{1}{2}}{x_1=x_2}$ to the example above allows stuck system states on $x_1=x_2$ to continuously progress along the line (illustrated below, left).
An alternative is \emph{hysteresis switching}~\citep{DBLP:books/sp/Liberzon03} which enlarges domains adjacent to the sliding mode so that a system that reaches the sliding surface is allowed to briefly continue following its current dynamics before switching.
For example, for a fixed $\varepsilon > 0$, the enlarged domains $\ivr_A \mnodefequiv x_1\geq x_2 - \varepsilon$ and $\ivr_B \mnodefequiv x_1 \leq x_2 + \varepsilon$ allows the stuck states to evolve off the line for a short distance.
This yields arbitrary switching in the overlapped part of both domains (illustrated below, right).
For a family of domains $\ivr_p$, $p \in \sigfam$ meeting conditions \circled{1} and \circled{2}, hysteresis switching can be introduced by replacing each $\ivr_p$ with its closed $\varepsilon$-neighborhood for some chosen $\varepsilon > 0$.

\begin{minipage}{0.22\textwidth}
\includegraphics[width=0.85\columnwidth,clip,trim=5 5 15 10]{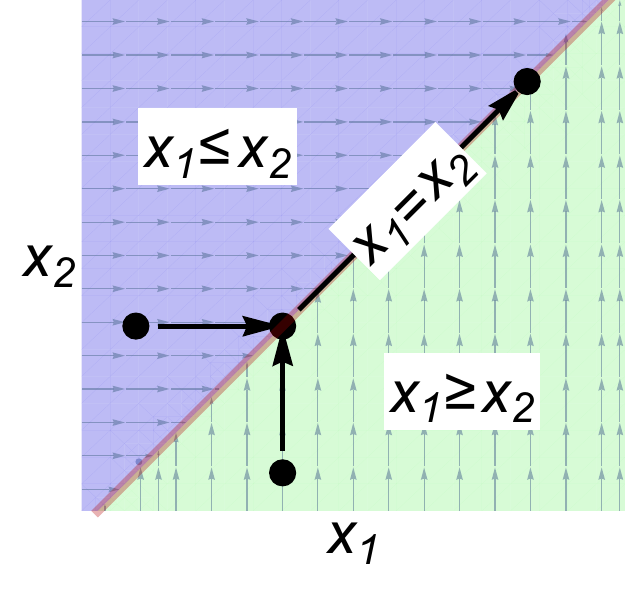}
\end{minipage}
\hfill\begin{minipage}{0.22\textwidth}
\includegraphics[width=0.85\columnwidth,clip,trim=5 5 15 10]{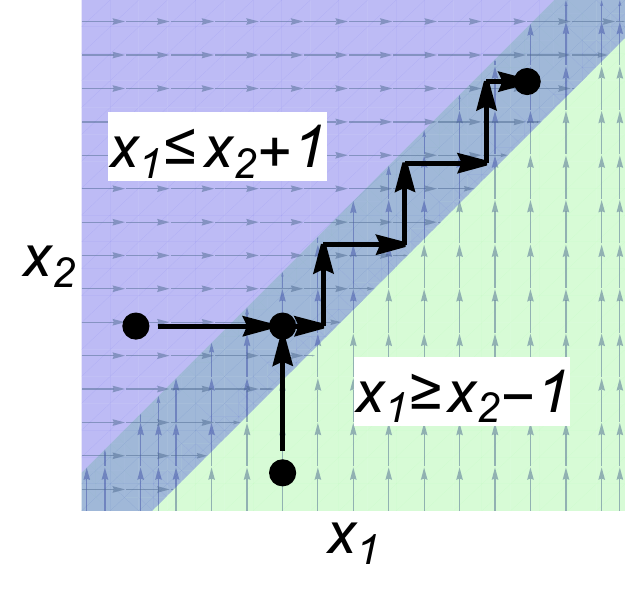}
\end{minipage}

To guarantee the absence of stuck states, by~\rref{thm:localprogress}, it suffices to decide validity of the formula \circled{3}: $\lorfold_{p \in \sigfam}{\sigliesai{f_p}{(\ivr_p)}}$, i.e., every point in the state space can switch to an ODE which locally progresses in its associated domain.
Models meeting conditions \circled{2} and \circled{3} also meet condition \circled{1}.

\subsubsection{Zeno behavior.}
Hybrid and switched system models can also exhibit \emph{Zeno behavior}, where the model makes infinitely many discrete transitions in a finite time interval~\citep{10.1002/rnc.592}.
Such behaviors are an artifact of the model and are not reflective of the real world.
Zeno traces are typically excluded when reasoning about hybrid system models~\citep{10.1002/rnc.592}, e.g.,~\rref{prop:adequacystateswitching} specifies safety for all \emph{finite} (thus non-Zeno) executions of state-dependent switching.
The detection of Zeno behavior in switched systems is left out of scope for this paper.

\section{Time-Dependent and Controlled Switching}
\label{sec:timectrlswitching}

\subsection{Time-Dependent Switching}
\label{sec:timeswitching}

The \emph{time-dependent switching} paradigm imposes timing constraints on switching signals.
To specify such constraints syntactically, each ODE in the family $p \in \sigfam$ is extended with a common, fresh clock variable $t$ with $\D{t}=1$ yielding ODEs of the form $\D{x}=f_p{(x)}, \D{t}=1$, and a fresh (discrete) flag variable $u$ is used to select and track the ODE to follow at each time.
One form of timing constraint is \emph{slow switching}, where the system switches arbitrarily between ODEs but must spend a minimum \emph{dwell time} $\tau > 0$ between each switch.
Sufficiently large dwell times can be used to stabilize some systems (see~\rref{sec:stability}).
Slow switching is modeled by the following hybrid program:
\begin{align*}
\slowswitch &\mnodefequiv \alpha_r ; \Big(\ifthen{t \geq \tau}{\alpha_r} ; \bigcup_{p \in \sigfam}{\big( \ptest{u {=} p} ; \pevolve{\D{x}{=}f_p{(x)}, \D{t}{=}1} \big)}\Big)^*\\
\alpha_r &\mnodefequiv~\pumod{t}{0}; \bigcup_{p \in \sigfam}{ \pumod{u}{p}}
\end{align*}

The program $\alpha_r$ resets the clock $t$ to $0$ and nondeterministically chooses a new value for the flag $u$.
For each loop iteration of $\slowswitch$, the guard $t \geq \tau$ checks if the current ODE has executed for at least time $\tau$ before running $\alpha_r$ to pick a new value for $u$.
The subsequent choice selects the ODE to follow based on the value of flag $u$.

\begin{prop}
A state is reachable by hybrid program $\slowswitch$ iff it is reachable in finite time by a switched system $\D{x}=f_p(x)$ for $p \in \sigfam$ following a switching signal $\sigma$ that spends at least time $\tau$ between its switching times.
\label{prop:adequacyslowswitching}
\end{prop}

\begin{thm}
From input ODEs $\pevolve{\D{x}=f_p{(x)}}$, $p \in \sigfam$ and formula $\rifvar$, there is a computable formula of real arithmetic $\phi$ such that formula $\rifvar$ is invariant for $\slowswitch$ iff $\phi$ is valid.
In particular, invariance for $\slowswitch$ is decidable.
\label{thm:slowdecidability}
\end{thm}

\subsection{Controlled Switching}
\label{sec:ctrlswitching}

The discrete fragment of hybrid programs can be used to flexibly model (computable) \emph{controlled switching} mechanisms, e.g., those that combine state-dependent and time-dependent switching constraints, or make complex switching decisions based on the state of the system.
An abstract controlled switching model is shown below, where program $\alpha_i$ initializes the system state (e.g., of the clock or flag) and $\alpha_u$ models a controller that assigns a decision $\pumod{u}{p}$.
\begin{align*}
\ctrlswitch \mnodefequiv \alpha_i ; \Big(
\alpha_u ; \bigcup_{p \in \sigfam}{\big( \ptest{u = p} ; \pevolvein{\D{x}=f_p{(x)}, \D{t}=1}{\ivr_p} \big)}
\Big)^*
\label{eqn:ctrlswitching}
\end{align*}

Hybrid program $\ctrlswitch$ resembles the shape of standard models of event-triggered and time-triggered systems in \dL~\citep{DBLP:books/sp/Platzer18} but is adapted for controlled switching.
The controller program $\alpha_u$ inspects the current state variables $x$ and the clock $t$.
It can modify the clock, e.g., by resetting it with $\pumod{t}{0}$, but $\alpha_u$ must not discretely change the state variables $x$.
The subsequent choice selects the ODE to follow based on the value of flag $u$ assigned in $\alpha_u$.

The slow switching model $\slowswitch$ is an instance of $\ctrlswitch$ where the controller program switches only after the dwell time is exceeded.
Another example is \emph{periodic switching}, where the controller periodically cycles through a family of ODEs.
Switching with sufficiently fast period can be used to stabilize a family of unstable ODEs, e.g., for linear ODEs whose system matrices have a stable convex combination~\citep{10.1080/00207728708964001}.
Without loss of generality, assume that $\sigfam\mnodefequiv \{1,\dots,m\}$, the desired switching order is $1, \dots, m$, and the periodic signal is required to follow the $i$-th ODE for exactly time $\zeta_i > 0$.
Periodic fast switching is modeled as an instance of $\ctrlswitch$ as follows:
\begin{align*}
\fastswitch &\mnodefequiv \ctrlswitch~\text{where}~\alpha_i \mnodefequiv~\pumod{t}{0};\pumod{u}{1}, \ivr_p \mnodefequiv t \leq \zeta_p,~\text{and} \\
\alpha_u &\mnodefequiv \bigcup_{p \in \sigfam} \texttt{if} (u = p \land t = \zeta_p) {\Bigg\{\begin{array}{l}
\pumod{t}{0} ; \pumod{u}{u+1}; \\
\ifthen{u > m}{\pumod{u}{1}}
\end{array}\Bigg\}}
\end{align*}

The system is initialized with $t=0$, $u = 1$ at the start of the cycle.
The controller program $\alpha_u$ then deterministically cycles through $u=1,\dots,m$ by discretely incrementing the flag variable whenever the time limit $\zeta_p$ for the currently chosen ODE is reached.
The domain constraints $\ivr_p$ respectively limit each ODE to run for at most time $\zeta_p$ as prescribed for the switched system.
\begin{prop}
A state is reachable by hybrid program $\fastswitch$ iff it is reachable in finite time by a switched system $\D{x}=f_p(x)$ for $p \in \{1,\dots,m\}$ following the switching signal $\sigma$ that periodically switches in the order $1,\dots,m$ according to the times $\zeta_1,\zeta_2,\dots,\zeta_m$ respectively.
\label{prop:adequacyfastswitching}
\end{prop}

A subtlety occurs in $\fastswitch$ and~\rref{prop:adequacyfastswitching} when one of the constituent ODEs exhibits \emph{finite time blowup} before reaching its switching time, e.g., consider switching between ODEs $\D{x}=1$ and $\D{x}=x^2$ with times $\zeta_1=\zeta_2=1$ starting from a state where $x=0$; the latter ODE blows up in the first cycle.
Mathematically, the switching signal $\sigma$ is simply ignored after the blowup time, but such blowup phenomena may not accurately reflect real world behavior.
Global existence of solutions for all ODEs in the switched system can be verified in \dL~\citep{DBLP:journals/fac/TanP}.

\section{Stability Verification in \KeYmaeraX}
\label{sec:stability}

This section shows how stability can be formally verified in \dL using the \KeYmaeraX theorem prover\footnote{All examples are formalized in \KeYmaeraX 4.9.2 at: \url{https://github.com/LS-Lab/KeYmaeraX-projects/blob/master/stability/switchedsystems.kyx}.} \citep{DBLP:conf/cade/FultonMQVP15} for the switched systems modeled by $\alpha \in \{ \arbswitch, \stateswitch, \slowswitch\}$.
For these systems, the origin $0 \in \reals^n$ is \emph{stable} iff the following formula is valid:
\begin{equation*}
\lforall{\varepsilon>0}{\lexists{\delta>0}{\lforall{x}{( \norm{x}^2 < \delta^2 \limply \dbox{\alpha}{\,\norm{x}^2 < \varepsilon^2}) } } }
\end{equation*}

This formula expresses that, for initial states sufficiently close to the origin ($\norm{x}^2 < \delta^2$ for $\delta > 0$), all states reached by hybrid program $\alpha$ from those states remain close to the origin ($\norm{x}^2 < \varepsilon^2$ for $\varepsilon > 0$).
By Propositions~\ref{prop:adequacyarbswitching},~\ref{prop:adequacystateswitching}, and~\ref{prop:adequacyslowswitching}, the formula specifies stability for the switched systems modeled by $\alpha\in \{ \arbswitch, \stateswitch, \slowswitch\}$ uniformly in their respective sets of switching signals~\citep{DBLP:books/sp/Liberzon03}.

Unlike invariance, a switched system can be stable (resp. unstable) even if all of its constituent ODEs are unstable (resp. stable), depending on the switching mechanism~\citep{DBLP:books/sp/Liberzon03}.
Stability verification for such systems is important because it provides formal guarantees that specific switching designs correctly eliminate potential instabilities in systems of interest.
An important technique for proving stability for ODEs and switched systems is to design an appropriate \emph{Lyapunov function}, i.e., an auxiliary energy measure that is non-increasing along all system trajectories~\citep{Liapounoff1907,DBLP:books/sp/Liberzon03}.

\begin{exmp}\label{ex:arbswitching}
Consider arbitrary switching $\arbswitch$ with ODEs:
\begin{align*}
&\pevolve{\D{x_1}= - x_1 + x_2^3, \D{x_2} = -x_1 - x_2} \\
&\pevolve{\D{x_1}= - x_1, \D{x_2} = -x_2}
\end{align*}

Both ODEs are stable and share the common Lyapunov function $\lterm \mnodefeq \frac{x_1^2}{2}+\frac{x_2^4}{4}$.
To prove stability for this example, the key idea is to show that $\lterm < k \land x_1^2+x_2^2 < \varepsilon$ is a loop invariant of $\arbswitch$, where $k$ is an upper bound on the initial value of $\lterm$ close to the origin.
\end{exmp}

\begin{exmp}\label{ex:stateswitching}
The following ODEs \circled{A} and \circled{B} are individually stable~\cite[Example 3.1]{DBLP:books/sp/Liberzon03}.
However, as illustrated below on the right, there is a switching signal that causes the system to diverge from the origin, i.e., these ODEs are \emph{not} stable under arbitrary switching.

\begin{minipage}{0.262\textwidth}
\begin{align*}
&\underbrace{\pevolve{\D{x_1}= -\frac{x_1}{8} - x_2, \D{x_2} = 2x_1 - \frac{x_2}{8}}}_{\hphantom{~\text{(solid blue)}}\circled{A}~\text{(solid blue)}} \\
&\underbrace{\pevolve{\D{x_1}= -\frac{x_1}{8} - 2x_2, \D{x_2} = x_1 - \frac{x_2}{8}}}_{\hphantom{~\text{(dashed red)}}\circled{B}~\text{(dashed red)}}
\end{align*}
\end{minipage}
\hfill\begin{minipage}{0.22\textwidth}~\quad
\includegraphics[width=0.85\columnwidth,clip,trim=5 5 15 10]{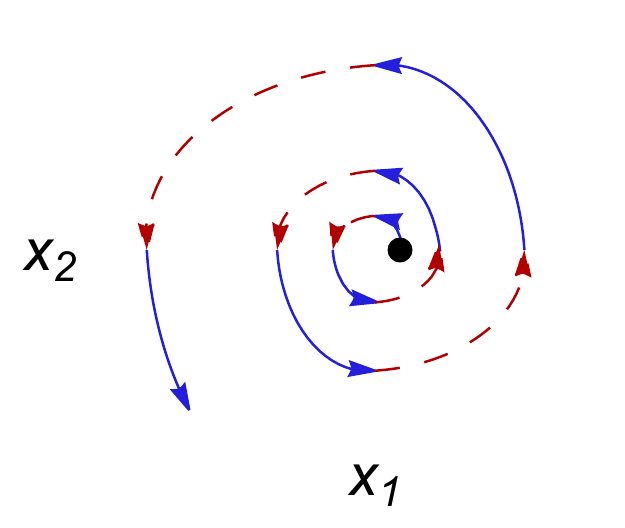}
\end{minipage}

Stability can be achieved by a state-dependent switching design with domains: \circled{A} $x_1x_2 \leq 0$ and \circled{B} $x_1x_2 \geq 0$.
The resulting system modeled by $\stateswitch$ has the common Lyapunov function $\lterm \mnodefeq x_1^2+x_2^2$.
The proof uses a loop invariant similar to~\rref{ex:arbswitching} and, crucially, checks the arithmetical Lyapunov function conditions for the derivative of $\lterm$ only on the respective domains for each ODE.
\end{exmp}

\begin{exmp}\label{ex:slowswitching}
The example ODEs \circled{A}, \circled{B} can also be stabilized by sufficiently slow switching in $\slowswitch$ with minimum dwell time $\tau \mnodefeq 3$ (the value of $\tau$ can be further optimized).
Here, two different Lyapunov functions are used: \circled{A} $2x_1^2 + x_2^2$ and \circled{B} $x_1^2 + 2x_2^2$.
The key proof idea is to bound both Lyapunov functions by decaying exponentials, and show that the dwell time $\tau$ is sufficiently large to ensure that both Lyapunov functions have decayed by an appropriate fraction when a switch occurs at time $t \geq \tau$.

The minimum dwell time principle can be used more generally to stabilize any family of stable linear ODEs~\citep{DBLP:books/sp/Liberzon03}.
For example, the ODE \circled{C} $ \D{x_1} = -x_1, \D{x_2} = -x_2$ is also stable and has the Lyapunov function $x_1^2+x_2^2$.
All three ODEs \circled{A}, \circled{B}, \circled{C} can be stabilized with the same dwell time $\tau \mnodefeq 3$.
The \KeYmaeraX proof required minimal changes, e.g., the loop invariants were updated to account for the new ODE \circled{C} and its Lyapunov function.
\end{exmp}

\section{Related Work}
\label{sec:relatedwork}

There are numerous hybrid system formalisms in the literature~\citep{10.2307/j.ctt7zvqf0, DBLP:books/sp/Liberzon03,SunGe, 4806347,10.2307/j.ctt7s02z, DBLP:conf/lics/Henzinger96,DBLP:journals/tcs/RonkkoRS03,DBLP:conf/aplas/LiuLQZZZZ10,DBLP:books/daglib/0025392,DBLP:books/sp/Platzer18}; see the cited articles and textbooks for further references.

Connections between several formalisms have been examined in prior work.
\cite{DBLP:books/daglib/0025392} shows how hybrid automata can be embedded into hybrid programs for their safety verification; the book also generalizes \dL with (disjunctive) differential-algebraic constraints that can be used to model and verify continuous dynamics with state-dependent switching~\cite[Chapter 3]{DBLP:books/daglib/0025392}.
This paper instead models switching with discrete program operators which enables compositional reasoning for the hybrid dynamics in switched systems.
\cite{DBLP:journals/tecs/SogokonGJ17} study hybrid automata models for ODEs with piecewise continuous right-hand sides and highlight various subtleties in the resulting models; similar subtleties for state-dependent switching models are presented in~\rref{sec:modelsubtleties}.
\cite{4806347,10.2307/j.ctt7s02z} show how impulsive differential equations, hybrid automata, and switched systems can all be understood as hybrid time models, and derive their properties using this connection; Theorems~\ref{thm:decidability} and \ref{thm:slowdecidability} are proved for switched systems using their hybrid program models.

\section{Conclusion}
This paper provides a blueprint for developing and verifying hybrid program models of switched systems.
These contributions enable several future directions, including: \begin{inparaenum}[\it i)]
\item formalizing \emph{asymptotic stability} for switched systems~\citep{DBLP:books/sp/Liberzon03,SunGe}, i.e., the systems are stable (\rref{sec:stability}) \emph{and} their trajectories tend to the origin over time;
\item modeling switched systems under more general continuous dynamics, e.g., differential inclusions~\citep{10.2307/j.ctt7s02z} or differential-algebraic constraints~\citep{DBLP:books/daglib/0025392};
\item developing practical proof automation for switched systems in \KeYmaeraX, e.g., automated synthesis and verification of invariants and Lyapunov functions for various switching mechanisms.
\end{inparaenum}

\paragraph*{Acknowledgments.} We thank the ADHS'21 anonymous reviewers for their helpful feedback on this paper.

\iflongversion
\bibliography{root-long}             %
\else                                %
\bibliography{root}             %
\fi

\iflongversion
\appendix
\section{Proofs}    %
\label{app:proofs}

This appendix provides full definitions and proofs for the results presented in the main paper.
Additional background material elided from~\rref{sec:background} is provided below for use in the proofs.

A \dL state $\iget[state]{\I} : \allvars \to \reals$ assigns a real value to each variable in $\allvars$.
The set of all variables $\allvars$ consists of the variables $x=(x_1,\dots,x_n)$ used to model the continuously evolving state of a switched system, and additional variables $\allvars \setminus \{x\}$ used as program auxiliaries in models, e.g., variables $u$ and $t$ in $\ctrlswitch$.
This paper focuses on the projection of \dL states on the variables $x$ so the (projected) \dL states $\iget[state]{\I}$ are equivalently treated as points in $\reals^n$.
Accordingly, the set of states where formula $\ivr$ is true is the set $\imodel{\I}{\ivr} \subseteq \reals^n$, and the transition relation for hybrid program $\alpha$ is $\iaccess[\alpha]{\I} \subseteq \reals^n \times \reals^n$ where $\iaccessible[\alpha]{\I}{\It}$ iff state $\iget[state]{\It} \in \reals^n$ is reachable from state $\iget[state]{\I} \in \reals^n$ by following $\alpha$.
The semantics of program auxiliaries is as usual~\citep{DBLP:books/sp/Platzer18}.

Switching signals $\sigma : [0,\infty) \to \sigfam$ are assumed to be \emph{well-defined}, i.e., $\sigma$ has finitely many discontinuities on each finite time interval in its domain $[0,\infty)$.
For finite $\sigfam$, this means $\sigma$ is a piecewise constant function with finitely many pieces on each finite time interval; intuitively, $\sigma$ prescribes a switching choice $p \in \sigfam$ on each piece.
For simplicity, $\sigma$ is also assumed to be right-continuous~\citep{10.2307/j.ctt7s02z}.
With these assumptions, switching signals are equivalently defined by a sequence of \emph{switching times} $0 = \tau_0 < \tau_1 < \tau_2 < \dots$ with $\tau_i \to \infty$ and a sequence $p_1,p_2, \dots \in \sigfam$ which specifies the values taken by $\sigma$ on each time interval:
\begin{equation}
\sigma(t) = \begin{cases}
p_1 & \text{if}~ \tau_0 \leq t < \tau_1 \\
p_2 & \text{if}~ \tau_1 \leq t < \tau_2 \\
& \cdots \\
p_i & \text{if}~ \tau_{i-1} \leq t < \tau_i
\end{cases}
\label{eqn:switching}
\end{equation}

For a switching signal $\sigma$ and initial state $\iget[state]{\I} \in \reals^n$, the solution $\solvar$ of the switched system is the function generated inductively on the sequences $\tau_i$ and $p_i$ as follows.
Define $\solvar(0) = \iget[state]{\I}$.
For switching time $\tau_i$ with $i \geq 1$, if $\solvar$ is defined at time $\tau_{i-1}$, then the definition of $\solvar$ is extended by considering the unique, right-maximal solution to the ODE $\D{x}=f_{p_i}(x)$ starting from $\solvar(\tau_{i-1})$~\citep{Chicone2006}, i.e., $\solvarr_i : [0,\zeta_i) \to \reals^n$ with $\solvarr_i(0) = \solvar(\tau_{i-1})$, $\D[t]{\solvarr_i(t)} = f_{p_i} (\solvarr_i(t))$, and $0 < \zeta_i \leq \infty$.
If $\zeta_i \leq \tau_i - \tau_{i-1}$, then the system \emph{blows up} before reaching the next switching time $\tau_i$, so define $\solvar(\tau_{i-1} + t) = \solvarr_i(t)$ on the bounded time interval $t \in [0,\zeta_i)$.
Otherwise, $\zeta_i > \tau_i - \tau_{i-1}$, then define $\solvar(\tau_{i-1}+t) = \solvarr_i(t)$ on the time interval $t \in [0,\tau_i - \tau_{i-1}]$.
This inductive construction uniquely defines a solution $\solvar : [0,\zeta) \to \reals^n$ associated with $\iget[state]{\I}$ and $\sigma$ for (right-maximal) time $\zeta > 0$.

The switched system \emph{reaches} $\solvar(t)$ at time $t \in [0,\zeta)$.
When the system is associated with a family of domains $\ivr_p$, $p \in \sigfam$, the switched system reaches $\solvar(t)$ while \emph{obeying} the domains iff for all $i \geq 1$ and time $\gamma \in [\tau_{i-1},\tau_{i}] \cap [0,t]$, the state $\solvar(\gamma)$ satisfies $\ivr_{p_i}$.

The \dL proof calculus used in the proofs of~\rref{lem:compinvariance} and~\rref{thm:slowdecidability} is briefly recalled here, a more comprehensive introduction is available elsewhere~\citep{DBLP:journals/jar/Platzer17,DBLP:books/sp/Platzer18}.
All derivations are presented in a classical sequent calculus with the usual rules for manipulating logical connectives and sequents such as~\irref{andl+allr}.
The semantics of \emph{sequent} \(\lsequent{\Gamma}{\fvarA}\) is equivalent to the formula \((\landfold_{\fvarB \in\Gamma} \fvarB) \limply \fvarA\) and a sequent is \emph{valid} iff its corresponding formula is valid.
Completed branches in a sequent proof are marked with $\lclose$.
An axiom (schema) is \emph{sound} iff all of its instances are valid.
A proof rule is \emph{sound} iff validity of all premises (above the rule bar) entails validity of the conclusion (below the rule bar).
Axioms and proof rules are \emph{derivable} if they can be deduced from sound \dL axioms and proof rules.
Soundness of the \dL axiomatization ensures that derived axioms and proof rules are sound~\citep{DBLP:journals/jar/Platzer17,DBLP:books/sp/Platzer18}.
The following axioms and proof rules of \dL are used in the proofs.

\begin{calculuscollection}
\begin{calculus}
\cinferenceRule[assignb|$\dibox{:=}$]{assignment / substitution axiom}
      {\linferenceRule[equiv]
        {\rfvar(e)}
        {\dbox{\pupdate{\umod{x}{\etermA}}}{\rfvar(x)}}
      }
      {\text{$\etermA$ free for $x$ in $\rfvar$}}
\end{calculus}\\
\begin{calculus}
\cinferenceRule[testb|$\dibox{?}$]{test}
{\linferenceRule[equiv]
  {(\ivr \limply \rfvar)}
  {\dbox{\ptest{\ivr}}{\rfvar}}
}{}

\cinferenceRule[composeb|$\dibox{{;}}$]{composition} %
      {\linferenceRule[equiv]
        {\dbox{\alpha}{\dbox{\beta}{\rfvar}}}
        {\dbox{\alpha;\beta}{\rfvar}}
      }{}

\end{calculus}\quad
\begin{calculus}
\cinferenceRule[choiceb|$\dibox{\cup}$]{axiom of nondeterministic choice}
{\linferenceRule[equiv]
  {\dbox{\alpha}{\rfvar} \land \dbox{\beta}{\rfvar}}
  {\dbox{\pchoice{\alpha}{\beta}}{\rfvar}}
}{}
\cinferenceRule[iterateb|$\dibox{{}^*}$]{iteration/repeat unwind} %
{\linferenceRule[equiv]
  {\rfvar \land \dbox{\alpha}{\dbox{\prepeat{\alpha}}{\rfvar}}}
  {\dbox{\prepeat{\alpha}}{\rfvar}}
}{}
\end{calculus}\\
\begin{calculus}
\dinferenceRule[loop|loop]{}
{\linferenceRule
  {\lsequent{\rfvar}{\dbox{\alpha}{\rfvar}}}
  {\lsequent{\rfvar}{\dbox{\prepeat{\alpha}}{\rfvar}}}
}{}
\end{calculus}\quad
\begin{calculus}
\cinferenceRule[G|G]{$\dbox{}{}$ generalization} %
{\linferenceRule[formula]
  {\lsequent{}{\rfvar}}
  {\lsequent{\Gamma}{\dbox{\alpha}{\rfvar}}}
}{}
\end{calculus}\quad
\begin{calculus}
\dinferenceRule[Mb|M${\dibox{\cdot}}$]{}
{\linferenceRule
  {\lsequent{\rrfvar}{\rfvar} \qquad \lsequent{\Gamma}{\dbox{\alpha}{\rrfvar}}}
  {\lsequent{\Gamma}{\dbox{\alpha}{\rfvar}}}
}{}
\end{calculus}\\
\begin{calculus}
\cinferenceRule[DGt|DGt]{differential ghost time}
{\linferenceRule[equiv]
  {\dbox{\pevolvein{\D{x}{=}\genDE{x},\D{t}{=}1}{\ivr(x)}}{\rfvar(x)}}
  {\dbox{\pevolvein{\D{x}{=}\genDE{x}}{\ivr(x)}}{\rfvar(x)}}
}{}
\end{calculus}
\end{calculuscollection}

\irlabel{qear|\usebox{\Rval}}
\irlabel{notr|$\lnot$\rightrule}
\irlabel{notl|$\lnot$\leftrule}
\irlabel{orr|$\lor$\rightrule}
\irlabel{orl|$\lor$\leftrule}
\irlabel{andr|$\land$\rightrule}
\irlabel{andl|$\land$\leftrule}
\irlabel{implyr|$\limply$\rightrule}
\irlabel{implyl|$\limply$\leftrule}
\irlabel{equivr|$\lbisubjunct$\rightrule}
\irlabel{equivl|$\lbisubjunct$\leftrule}
\irlabel{id|id}
\irlabel{cut|cut}
\irlabel{weakenr|W\rightrule}
\irlabel{weakenl|W\leftrule}
\irlabel{existsr|$\exists$\rightrule}
\irlabel{existsrinst|$\exists$\rightrule}
\irlabel{alll|$\forall$\leftrule}
\irlabel{alllinst|$\forall$\leftrule}
\irlabel{allr|$\forall$\rightrule}
\irlabel{existsl|$\exists$\leftrule}
\irlabel{iallr|i$\forall$}
\irlabel{iexistsr|i$\exists$}

Axioms~\irref{assignb+testb+composeb+choiceb+iterateb} unfold box modalities of their respective hybrid programs according to their semantics.
Rule~\irref{loop} is the loop induction rule, rule~\irref{G} is G\"odel generalization, and rule~\irref{Mb} is the derived monotonicity rule for box modality postconditions; antecedents that have no free variables bound in $\alpha$ are soundly kept across uses of rules~\irref{loop+G+Mb}~\citep{DBLP:journals/jar/Platzer17,DBLP:books/sp/Platzer18}.
Axiom~\irref{DGt} is an instance of the more general~\emph{differential ghosts} axiom of \dL, which adds (or removes) a fresh linear system of ODEs to an ODE $\D{x}=\genDE{x}$ for the sake of the proof.

\begin{pf*}{Proof of~\rref{prop:adequacyarbswitching}.}
This follows from~\rref{prop:adequacystateswitching} with $\ivr_p \mnodefequiv \ltrue$ for all $p\in\sigfam$. $\hfill\qed$
\end{pf*}

\begin{pf*}{Proof of~\rref{prop:adequacystateswitching}.}
Both directions of the proposition are proved separately for an initial state $\iget[state]{\I} \in \reals^n$.

``$\Rightarrow$''. Suppose $\iaccessible[\stateswitch]{\I}{\It}$.
By the semantics of \dL loops, there is a sequence of states $\iget[state]{\I} = \iget[state]{\I}_0, \iget[state]{\I}_1, \dots, \iget[state]{\I}_n = \iget[state]{\It}$ for some $n \geq 0$ and for each $1 \leq i \leq n$, the states transition according to $(\iget[state]{\I}_{i-1},\iget[state]{\I}_{i}) \in \iaccess[\bigcup_{p \in \sigfam}{ \pevolvein{\D{x}=f_p{(x)}}{\ivr_p}}]{\I}$.
In particular, for each $1 \leq i \leq n$, there is a choice $p_i$ where state $\iget[state]{\I}_{i-1}$ reaches $\iget[state]{\I}_{i}$ by evolving according to the ODE $\D{x}=f_{p_i}{(x)}$ for some time $\zeta_i \geq 0$ and staying within the domain $\ivr_{p_i}$ for all times $0 \leq t \leq \zeta_i$ during its evolution.

The finite sequences $(\iget[state]{\I}_0,\iget[state]{\I}_1,\dots,\iget[state]{\I}_n)$, $(\zeta_1,\dots,\zeta_n)$ and $(p_1,\dots,p_n)$ correspond to a well-defined switching signal as follows.
First, remove from all sequences the indexes $1 \leq i \leq n$ with $\zeta_i=0$.
This yields new sequences $(\tilde{\iget[state]{\I}}_0, \tilde{\iget[state]{\I}}_1,\dots,\tilde{\iget[state]{\I}}_m)$, $(\tilde{\zeta}_1,\dots,\tilde{\zeta}_m)$, and $(\tilde{p}_1,\dots,\tilde{p}_m)$ where ${\tilde{\zeta}_i > 0}$.
Consider the switching signal $\sigma$ with switching times $\tau_i = \sum_{j=1}^{i} \tilde{\zeta}_j$ for $1 \leq i < m$ and $\tau_{i} = \tau_{i-1} + 1$ for $i \geq m$, so $\tau_1 < \tau_2 < \dots$ and $\tau_i \to \infty$.
Furthermore, extend the sequence of switching choices with $\tilde{p}_i = \tilde{p}_m$ for $i > m$.
By construction using~\rref{eqn:switching}, $\sigma$ is well-defined and the solution $\solvar$ associated with $\sigma$ from $\iget[state]{\I}$ reaches $\iget[state]{\It}$ at time $\sum_{j=1}^{m} \tilde{\zeta}_j$ and obeys the domains $\ivr_{\tilde{p}_i}$ until that time.

``$\Leftarrow$''. Let $\sigma$ be a switching signal and $\solvar : [0,\zeta) \to \reals^n$ be the associated switched system solution from $\iget[state]{\I}$.
Suppose that the switched system reaches $\solvar(t)$ for $t \in [0,\zeta)$ while obeying the domains $\ivr_p$.
To show $(\iget[state]{\I}, \solvar(t)) \in \iaccess[\stateswitch]{\I}$, by the semantics of \dL loops, it suffices to construct a sequence of states $\iget[state]{\I}=\iget[state]{\I}_0, \iget[state]{\I}_1, \dots, \iget[state]{\I}_n$ for some finite $n$, with $\iget[state]{\I}_n = \solvar(t)$, and $(\iget[state]{\I}_{i-1},\iget[state]{\I}_{i}) \in \iaccess[\bigcup_{p \in \sigfam}{ \pevolvein{\D{x}=f_p{(x)}}{\ivr_p}}]{\I}$ for $1 \leq i \leq n$.

By~\rref{eqn:switching}, $\sigma$ is equivalently defined by a sequence of switching times $\tau_0 < \tau_1 < \tau_2 < \dots$ and a sequence of switching choices $p_1,p_2, \dots$, where $p_i \in \sigfam$.
Let $\tau_{n}$ be the first switching time such that $t \leq \tau_{n}$; the index $n$ exists since $\tau_i \to \infty$.
Define the state sequence $\iget[state]{\I}_i \mnodefeq \solvar(\tau_i)$ for $0 \leq i < n$ and $\iget[state]{\I}_n = \solvar(t)$.
Note that $\iget[state]{\I}_0 = \iget[state]{\I}$ by definition of $\solvar(0)$.
It suffices to show $(\iget[state]{\I}_{i-1},\iget[state]{\I}_{i}) \in \iaccess[\pevolvein{\D{x}=f_{p_i}{(x)}}{\ivr_{p_i}}]{\I}$ for $1 \leq i \leq n$, but this follows by construction of $\solvar$ because $\iget[state]{\I}_{i}$ is reached from $\iget[state]{\I}_{i-1}$ by following the solution to ODE $\D{x}=f_{p_i}(x)$, and, by assumption, $\solvar(\gamma)$ satisfies $\ivr_{p_i}$ for $\gamma \in [\tau_{i-1},\tau_{i}] \cap [0,t]$.
$\hfill\qed$
\end{pf*}

\begin{pf*}{Proof of~\rref{lem:compinvariance}.}
The following axiom is syntactically derived in \dL. It syntactically expresses that invariance for $\stateswitch$ (left-hand side) is equivalent to invariance for all of its constituent ODEs (right-hand side).
\[
\dinferenceRule[sdsi|Inv$_{\text{state}}$]{state-dependent switching invariance}
{\linferenceRule[equivl]
  {
   \landfold_{p \in \sigfam}{\lforall{x}{(\rifvar \limply \dbox{\pevolvein{\D{x} = f_p(x)}{\ivr_p}}{\rifvar})}}
  }
  {\lforall{x}{(\rifvar \limply \dbox{\stateswitch}{\rifvar})}}
}{}
\]
Both directions of axiom~\irref{sdsi} are derived separately.
\begin{itemize}
\item[``$\lylpmi$''] The (easier) ``$\lylpmi$'' direction uses rule~\irref{loop} to prove that $\rifvar$ is a loop invariant of $\stateswitch$.
The antecedent is abbreviated $\Gamma \mnodefequiv \landfold_{p \in \sigfam}{\lforall{x}{(\rifvar \limply \dbox{\pevolvein{\D{x} = f_p(x)}{\ivr_p}}{\rifvar})}}$; $\Gamma$ is constant for $\stateswitch$, so it is soundly kept across the use of rule~\irref{loop}.
The subsequent~\irref{choiceb+andr} step unfolds the nondeterministic choice in $\stateswitch$'s loop body, yielding a premise for each ODE in $\sigfam$.
These premises are indexed by $p \in \sigfam$ below and are all proved propositionally from $\Gamma$.
{\renewcommand*{\arraystretch}{1.2}%
\begin{sequentdeduction}[array]
  \linfer[allr+implyr]{
  \linfer[loop]{
  \linfer[choiceb+andr]{
  \linfer[andl+alll+implyl]{
    \lclose
  }
    {\lsequent{\Gamma, \rifvar}{\dbox{\pevolvein{\D{x}=f_p{(x)}}{\ivr_p}}{\rifvar}}}
  }
    {\lsequent{\Gamma, \rifvar}{\dbox{ \bigcup_{p \in \sigfam}{ \pevolvein{\D{x}=f_p{(x)}}{\ivr_p}}}{\rifvar}}}
  }
    {\lsequent{\Gamma, \rifvar}{\dbox{\stateswitch}{\rifvar}}}
  }
  {\lsequent{\Gamma}{\lforall{x}{(\rifvar \limply \dbox{\stateswitch}{\rifvar})}}}
\end{sequentdeduction}}%

\item[``$\limply$''] The ``$\limply$'' direction shows that a run of ODE $\pevolvein{\D{x} = f_p(x)}{\ivr_p}$, $p \in \sigfam$ must also be a run of $\stateswitch$, so if formula $\rifvar$ is true for all runs of $\stateswitch$, it must also be true for all runs of the constituent ODEs.
The derivation starts by logical unfolding, with abbreviated antecedent $\Gamma \mnodefequiv \lforall{x}{(\rifvar \limply \dbox{\stateswitch}{\rifvar})}$; the resulting premises are indexed by $p \in \sigfam$ below.
{%
\begin{sequentdeduction}[array]
  \linfer[andr+allr+implyr]{
  \linfer[alll+implyl]{
    \lsequent{\dbox{\stateswitch}{\rifvar}}{\dbox{\pevolvein{\D{x} = f_p(x)}{\ivr_p}}{\rifvar}}
  }
    {\lsequent{\Gamma,\rifvar}{\dbox{\pevolvein{\D{x} = f_p(x)}{\ivr_p}}{\rifvar}}}
  }
  {\lsequent{\Gamma}{\landfold_{p \in \sigfam}{\lforall{x}{(\rifvar \limply \dbox{\pevolvein{\D{x} = f_p(x)}{\ivr_p}}{\rifvar})}}}}
\end{sequentdeduction}}%
Next, axiom~\irref{iterateb} unfolds the loop in the antecedent before axiom~\irref{choiceb} chooses the branch corresponding to $p \in \sigfam$ in the loop body.
The loop body in $\stateswitch$ is abbreviated $\alpha_l \mnodefequiv \bigcup_{p \in \sigfam}{ \pevolvein{\D{x}=f_p{(x)}}{\ivr_p}}$ below.
{\renewcommand*{\arraystretch}{1.2}%
\begin{sequentdeduction}[array]
  \linfer[iterateb+andl]{
  \linfer[choiceb+andl]{
  \linfer[Mb]{
  \linfer[iterateb+andl]{
    \lclose
  }
    {\lsequent{\dbox{\stateswitch}{\rifvar}}{\rifvar}}
  }
    {\lsequent{\dbox{\pevolvein{\D{x}=f_p{(x)}}{\ivr_p}}{\dbox{\stateswitch}{\rifvar}}}{\dbox{\pevolvein{\D{x} = f_p(x)}{\ivr_p}}{\rifvar}}}
  }
    {\lsequent{\dbox{\alpha_l}{\dbox{\stateswitch}{\rifvar}}}{\dbox{\pevolvein{\D{x} = f_p(x)}{\ivr_p}}{\rifvar}}}
  }
    {\lsequent{\dbox{\stateswitch}{\rifvar}}{\dbox{\pevolvein{\D{x} = f_p(x)}{\ivr_p}}{\rifvar}}}
\end{sequentdeduction}}%
The derivation is completed using rule~\irref{Mb} to monotonically strengthen the postcondition, then unfolding the resulting antecedent with axiom~\irref{iterateb}. $\hfill\qed$
\end{itemize}
\end{pf*}

\begin{pf*}{Proof of~\rref{thm:decidability}.}
Recall for input ODE $\D{x}=\genDE{x}$ and formula of real arithmetic $\ivr$, there is a computable formula of real arithmetic $\sigliesai{f}{(\ivr)}$ characterizing the states from which $\D{x}=f(x)$ locally progresses into $\ivr$ (similarly, formula $\sigliesai{-f}{(\ivr)}$ characterizes local exit from $\ivr$).
Unlike the earlier presentation~\citep{DBLP:journals/jacm/PlatzerT20}, this paper explicitly indicates the ODE dependency in formula $\sigliesai{f}{(\ivr)}$ for notational clarity when considering switched systems involving multiple different ODEs.

By~\citet[Theorem B.5]{DBLP:journals/jacm/PlatzerT20}, the following axiom is derivable in \dL for polynomial ODEs $\D{x}=\genDE{x}$ and real arithmetic formulas $\rfvar, \ivr$.
\[
\dinferenceRule[SAIQ|SAI{$\&$}]{Semianalytic invariant with domains axiom}
{\linferenceRule[equivl]
  {
  \left(\begin{array}{l}
  \lforall{x}{\big(\rfvar \land \ivr \land \sigliesai{f}{(\ivr)} \limply \sigliesai{f}{(\rfvar)}\big)} \land \\
  \lforall{x}{\big(\lnot{\rfvar} \land \ivr \land \sigliesai{-f}{(\ivr)} \limply \sigliesai{-f}{(\lnot{\rfvar})}\big)}
  \end{array}\right)
  }
  {\lforall{x}{(\rfvar \limply \dbox{\pevolvein{\D{x}=\genDE{x}}{\ivr}}{\rfvar})}}
}{}
\]

Chaining the equivalence~\irref{sdsi} from~\rref{lem:compinvariance} and~\irref{SAIQ} syntactically derives the following equivalence in \dL:
\[
\dinferenceRule[sdsiSAIQ|SAI$_{\text{state}}$]{state dependent switching invariance}
{\linferenceRule[equivl]
  {
   \landfold_{p \in \sigfam}{
 \left(\begin{array}{l}
  \lforall{x}{\big(\rifvar \land \ivr_p \land \sigliesai{f_p}{(\ivr_p)} \limply \sigliesai{f_p}{(\rifvar)}\big)} \land \\
  \lforall{x}{\big(\lnot{\rifvar} \land \ivr_p \land \sigliesai{-f_p}{(\ivr_p)} \limply \sigliesai{-f_p}{(\lnot{\rifvar})}\big)}
  \end{array}\right)
  }
  }
  {\lforall{x}{(\rifvar \limply \dbox{\stateswitch}{\rifvar})}}
}{}
\]

Derived axiom~\irref{sdsiSAIQ} equivalently characterizes invariance of formula $\rifvar$ for $\stateswitch$ by a decidable formula of first-order real arithmetic~\citep{Tarski} on its right-hand side. Therefore, invariance for state-dependent switched systems is decidable.
$\hfill\qed$
\end{pf*}

\begin{pf*}{Proof of~\rref{thm:localprogress}.}
Local progress is specified using \dL in~\citet[Section 5]{DBLP:journals/jacm/PlatzerT20} and characterized by a provably equivalent formula of real arithmetic in~\citet[Theorem 6.6]{DBLP:journals/jacm/PlatzerT20}. $\hfill\qed$
\end{pf*}

\begin{pf*}{Proof of~\rref{prop:adequacyslowswitching}.}
The proof is similar to~\rref{prop:adequacystateswitching} but with fresh auxiliary variables $t,u$ used to control the switching signal.
Let $\tau > 0$ be the dwell time constraint of the system.
Both directions of the proposition are proved separately for an initial state $\iget[state]{\I} \in \reals^n$.

``$\Rightarrow$''. Suppose $\iaccessible[\slowswitch]{\I}{\It}$.
The program $\alpha_r$ resets the clock $t$ to $0$ and sets the value of flag $u$ to $p \in \sigfam$, but leaves the state variables $x$ unchanged.
By the semantics of \dL programs, there is a sequence of states $\iget[state]{\I} = \iget[state]{\I}_0, \iget[state]{\I}_1, \dots, \iget[state]{\I}_n = \iget[state]{\It}$ for some $n \geq 0$ and for each $1 \leq i \leq n$, there is a choice $p_i$ where state $\iget[state]{\I}_{i-1}$ reaches $\iget[state]{\I}_{i}$ by following the ODE $\D{x}=f_{p_i}{(x)}$ for some time ${\zeta_i \geq 0}$.
Extract compacted sequences from $(\iget[state]{\I}_0,\iget[state]{\I}_1,\dots,\iget[state]{\I}_n)$, $(\zeta_1,\dots,\zeta_n)$ and $(p_1,\dots,p_n)$ as follows: while there is an index ${i \geq 1}$ such that $p_i = p_{i+1}$, replace $\zeta_i$ with $\zeta_{i} + \zeta_{i+1}$, $\iget[state]{\I}_i$ with $\iget[state]{\I}_{i+1}$ and delete the index $i+1$ from all sequences.
Intuitively, this compaction repeatedly combines adjacent runs of the loop body of $\slowswitch$ from the same ODE, yielding the sequences $(\tilde{\iget[state]{\I}}_0,\tilde{\iget[state]{\I}}_1,\dots,\tilde{\iget[state]{\I}}_m)$, $(\tilde{\zeta}_1,\dots,\tilde{\zeta}_m)$, and $(\tilde{p}_1,\dots,\tilde{p}_m)$ where $\tilde{\iget[state]{\I}}_0 = \iget[state]{\I}$, $\tilde{\iget[state]{\I}}_m = \iget[state]{\I}_n = \iget[state]{\It}$ and for $i \geq 1$, $\tilde{\iget[state]{\I}}_{i-1}$ reaches $\tilde{\iget[state]{\I}}_{i}$ following the ODE $\D{x}=f_{\tilde{p}_i}{(x)}$ by uniqueness of ODE solutions~\citep{Chicone2006}.
Furthermore, $\tilde{p}_i \not= \tilde{p}_{i-1}$ for $i \geq 1$ and $\tilde{\zeta}_i \geq \tau > 0$ for $1 \leq i < m$ because the guard $t \geq \tau$ in the loop body of $\slowswitch$ allows switching only when the dwell time $\tau$ has elapsed.

Consider the switching signal $\sigma$ with switching times $\tau_i = \sum_{j=1}^{i} \tilde{\zeta}_j$ for $1 \leq i < m$ and $\tau_{i} = \tau_{i-1} + \tau$ for $i \geq m$, so $\tau_i \to \infty$.
Note $\tau_{i}-\tau_{i-1} = \tilde{\zeta}_i \geq \tau$ for $i \geq 1$.
Furthermore, extend the sequence of switching choices with $\tilde{p}_i = \tilde{p}_m$ for $i > m$.
By construction using~\rref{eqn:switching}, $\sigma$ is well-defined, spends at least time $\tau$ between its switching times, and the solution $\solvar$ associated with $\sigma$ from $\iget[state]{\I}$ reaches $\iget[state]{\It}$ at time $\sum_{j=1}^{m} \tilde{\zeta}_j$.

``$\Leftarrow$''. Let $\sigma$ be a switching signal that spends at least time $\tau$ between its switching times and $\solvar : [0,\zeta) \to \reals^n$ be the associated switched system solution from $\iget[state]{\I}$.
Suppose the switched system reaches $\solvar(t)$ for $t \in [0,\zeta)$.
To show $(\iget[state]{\I}, \solvar(t)) \in \iaccess[\slowswitch]{\I}$, by the semantics of \dL programs, it suffices to construct a sequence of states $\iget[state]{\I}=\iget[state]{\I}_0, \iget[state]{\I}_1, \dots, \iget[state]{\I}_n$ for some finite $n$, with $\iget[state]{\I}_n = \solvar(t)$ and $\iget[state]{\I}_{i-1}$ reaches $\iget[state]{\I}_{i}$ by following the loop body of $\slowswitch$ for $1 \leq i \leq n$.

By~\rref{eqn:switching}, $\sigma$ is equivalently defined by a sequence of switching times $\tau_0, \tau_1, \dots$ with $\tau_{i}-\tau_{i-1} \geq \tau > 0$ for $i\geq 1$ and a sequence of switching choices $p_1, p_2, \dots$, where $p_i \in \sigfam$.
Let $\tau_{n}$ be the first switching time such that $t \leq \tau_{n}$; the index $n$ exists since $\tau_i \to \infty$.
Define the state sequence $\iget[state]{\I}_i \mnodefeq \solvar(\tau_i)$ for $0 \leq i < n$ and $\iget[state]{\I}_n = \solvar(t)$.
Note that $\iget[state]{\I}_0 = \iget[state]{\I}$ by definition of $\solvar(0)$.
By construction of $\solvar$, $\iget[state]{\I}_{i}$ is reached from $\iget[state]{\I}_{i-1}$ by following the solution to ODE $\D{x}=f_{p_i}(x)$.
Moreover, since the switching times satisfy $\tau_i - \tau_{i-1} \geq \tau$ for $1 \leq i < n$, the guard $t \geq \tau$ is satisfied for each run of the loop body of $\slowswitch$.
$\hfill\qed$
\end{pf*}

\begin{pf*}{Proof of~\rref{thm:slowdecidability}.}
Similar to~\rref{lem:compinvariance}, the following axiom will be syntactically derived in \dL, assuming the dwell time $\tau > 0$ is a positive constant.
\[
\dinferenceRule[tdsi|Inv$_{\text{slow}}$]{time-dependent switching invariance}
{\linferenceRule[equiv]
  {
   \landfold_{p \in \sigfam}{\lforall{x}{(\rifvar \limply \dbox{\pevolve{\D{x}=f_p(x)}}{\rifvar})}}
  }
  {\lforall{x}{(\rifvar \limply \dbox{\slowswitch}{\rifvar})}}
}{}
\]

Axiom~\irref{tdsi} says that invariance of formula $\rifvar$ for a slow switching system is equivalent to invariance of $\rifvar$ for each of its constituent ODEs.
The two directions of axiom~\irref{tdsi} are derived separately similar to the derivation of axiom~\irref{sdsi} but with additional steps to unfold the program $\alpha_r \mnodefequiv~\pumod{t}{0}; \bigcup_{p \in \sigfam}{ \pumod{u}{p}}$ and to handle the fresh variables $u,t$ it uses.
The loop body in $\slowswitch$ is abbreviated $\alpha_l \mnodefequiv \ifthen{t \geq \tau}{\alpha_r} ; \bigcup_{p \in \sigfam}{\big( \ptest{u = p} ; \pevolve{\D{x}=f_p{(x)}, \D{t}=1} \big)}$.

\begin{itemize}
\item[``$\lylpmi$''] The (easier) ``$\lylpmi$'' direction uses rule~\irref{loop} to prove that $\rifvar$ is a loop invariant of $\slowswitch$.
The antecedent is abbreviated $\Gamma \mnodefequiv \landfold_{p \in \sigfam}{\lforall{x}{(\rifvar \limply \dbox{\pevolve{\D{x} = f_p(x)}}{\rifvar})}}$.
The derivation is identical to the ``$\lylpmi$'' direction of~\irref{sdsi} except the use of axiom~\irref{composeb} and rule~\irref{G} to soundly skip over the discrete programs that set variables $u,t$.
Intuitively, \irref{composeb} and \irref{G} are used because invariance for $\slowswitch$ is independent of which (nondeterministic) choice of ODE is followed.
The antecedents $\Gamma,\rifvar$ are soundly kept across uses of rule~\irref{G} because they do not mention variables $u,t$.
In the penultimate step, axiom~\irref{DGt} removes the clock ODE $\D{t}=1$ and the derivation is completed with~\irref{andl+alll+implyl}.
Premises are indexed by $p \in \sigfam$ after the~\irref{choiceb+andr} step.

{\renewcommand*{\arraystretch}{1.1}%
\begin{sequentdeduction}[array]
  \linfer[allr+implyr]{
  \linfer[composeb+G]{
  \linfer[loop]{
  \linfer[composeb+G]{
  \linfer[choiceb+andr]{
  \linfer[composeb+G]{
  \linfer[DGt]{
  \linfer[andl+alll+implyl]{
    \lclose
  }
    {\lsequent{\Gamma, \rifvar}{\dbox{\pevolve{\D{x}=f_p{(x)}}}{\rifvar}}}
  }
    {\lsequent{\Gamma, \rifvar}{\dbox{\pevolve{\D{x}=f_p{(x)}, \D{t}=1}}{\rifvar}}}
  }
    {\lsequent{\Gamma, \rifvar}{\dbox{\ptest{u = p} ; \pevolve{\D{x}=f_p{(x)}, \D{t}=1}}{\rifvar}}}
  }
    {\lsequent{\Gamma, \rifvar}{\dbox{\bigcup_{p \in \sigfam}{\big( \ptest{u = p} ; \pevolve{\D{x}=f_p{(x)}, \D{t}=1} \big)}}{\rifvar}}}
  }
    {\lsequent{\Gamma, \rifvar}{\dbox{\alpha_l}{\rifvar}}}
  }
    {\lsequent{\Gamma, \rifvar}{\dbox{\alpha_l^*}{\rifvar}}}
  }
    {\lsequent{\Gamma, \rifvar}{\dbox{\slowswitch}{\rifvar}}}
  }
  {\lsequent{\Gamma}{\lforall{x}{(\rifvar \limply \dbox{\slowswitch}{\rifvar})}}}
\end{sequentdeduction}}%

\item[``$\limply$''] The ``$\limply$'' direction shows that a run of ODE $\pevolve{\D{x} = f_p(x)}$, $p \in \sigfam$ must also be a run of $\slowswitch$, so if formula $\rifvar$ is true for all runs of $\slowswitch$, it must also be true for all runs of the constituent ODEs.
The derivation starts by logical unfolding, with abbreviated antecedent $\Gamma \mnodefequiv \lforall{x}{(\rifvar \limply \dbox{\slowswitch}{\rifvar})}$.
Premises are indexed by $p \in \sigfam$.
{%
\begin{sequentdeduction}[array]
  \linfer[andr+allr+implyr]{
  \linfer[alll+implyl]{
    \lsequent{\dbox{\slowswitch}{\rifvar}}{\dbox{\pevolve{\D{x} = f_p(x)}}{\rifvar}}
  }
    {\lsequent{\Gamma,\rifvar}{\dbox{\pevolve{\D{x} = f_p(x)}}{\rifvar}}}
  }
  {\lsequent{\Gamma}{\landfold_{p \in \sigfam}{\lforall{x}{(\rifvar \limply \dbox{\pevolve{\D{x} = f_p(x)}}{\rifvar})}}}}
\end{sequentdeduction}}%
Next, axioms~\irref{composeb+assignb+choiceb} unfolds program $\alpha_r$ in $\slowswitch$, setting $t=0$ and choosing $p$ for the value of flag $u$.
Axiom~\irref{iterateb} unfolds the loop in the antecedents and the \texttt{if} program in $\alpha_l$ is skipped using axioms~\irref{choiceb+testb} because its guard formula $t \geq \tau$ contradicts the antecedent $t=0$.
This leaves the choice abbreviated $\alpha_c \mnodefequiv \bigcup_{p \in \sigfam}{\ptest{u = p} ; \pevolve{\D{x}=f_p{(x)}, \D{t}=1}}$, which is unfolded with axioms~\irref{choiceb+composeb+testb} according to the chosen value of flag $u$.
Axiom~\irref{DGt} then removes the clock ODE $\D{t}=1$ from the antecedent box modality.
{\renewcommand*{\arraystretch}{1.2}%
\begin{sequentdeduction}[array]
  \linfer[composeb+assignb+choiceb]{
  \linfer[iterateb+andl]{
  \linfer[choiceb+testb]{
  \linfer[choiceb+composeb+testb]{
  \linfer[DGt]{
  \linfer[Mb]{
  \linfer[iterateb+andl]{
    \lclose
  }
    {\lsequent{\dbox{\prepeat{\alpha_l}}{\rifvar}}{\rifvar}}
  }
    {\lsequent{\dbox{\pevolve{\D{x}=f_p{(x)}}}{\dbox{\prepeat{\alpha_l}}{\rifvar}}}{\dbox{\pevolve{\D{x} = f_p(x)}}{\rifvar}}}
  }
    {\lsequent{\dbox{\pevolve{\D{x}=f_p{(x)}, \D{t}=1}}{\dbox{\prepeat{\alpha_l}}{\rifvar}}}{\dbox{\pevolve{\D{x} = f_p(x)}}{\rifvar}}}
  }
    {\lsequent{u=p, \dbox{\alpha_c}{\dbox{\prepeat{\alpha_l}}{\rifvar}}}{\dbox{\pevolve{\D{x} = f_p(x)}}{\rifvar}}}
  }
    {\lsequent{t=0, u=p, \dbox{\alpha_l}{\dbox{\prepeat{\alpha_l}}{\rifvar}}}{\dbox{\pevolve{\D{x} = f_p(x)}}{\rifvar}}}
  }
    {\lsequent{t=0, u=p,\dbox{\prepeat{\alpha_l}}{\rifvar}}{\dbox{\pevolve{\D{x} = f_p(x)}}{\rifvar}}}
  }
    {\lsequent{\dbox{\slowswitch}{\rifvar}}{\dbox{\pevolve{\D{x} = f_p(x)}}{\rifvar}}}
\end{sequentdeduction}}%
The derivation is completed using rule~\irref{Mb} to monotonically strengthen the postcondition, then unfolding the resulting antecedent with axiom~\irref{iterateb}.
\end{itemize}

Chaining the equivalence~\irref{tdsi} and~\irref{SAIQ} (with formula $\ivr \mnodefequiv \ltrue$) derives the following equivalence in \dL:
\[
\dinferenceRule[tdsiSAIQ|SAI$_{\text{slow}}$]{time dependent switching invariance}
{\linferenceRule[equiv]
  {
   \landfold_{p \in \sigfam}{
 \left(\begin{array}{l}
  \lforall{x}{\big(\rifvar \limply \sigliesai{f_p}{(\rifvar)}\big)} \land \\
  \lforall{x}{\big(\lnot{\rifvar} \limply \sigliesai{-f_p}{(\lnot{\rifvar})}\big)}
  \end{array}\right)
  }
  }
  {\lforall{x}{(\rifvar \limply \dbox{\slowswitch}{\rifvar})}}
}{}
\]
Derived axiom~\irref{tdsiSAIQ} characterizes invariance for slow switching by a decidable formula of first-order real arithmetic~\citep{Tarski}. Thus, invariance for slow switching systems is decidable. $\hfill\qed$
\end{pf*}

\newcommand{\offmod}[1]{[#1]_m}

\begin{pf*}{Proof of~\rref{prop:adequacyfastswitching}.}
The proof is similar to Propositions~\ref{prop:adequacystateswitching} and~\ref{prop:adequacyslowswitching} with auxiliary fresh variables $t,u$ used to control the switching signal.
Let $\sigfam\mnodefequiv \{1,\dots,m\}$ with the switching order $1, \dots, m$, and where the periodic signal is required to follow the $i$-th ODE for exactly time $\zeta_i > 0$ for $i =1,\dots,m$.
Abbreviate $\offmod{i} \mnodefeq ((i - 1) \bmod m) + 1$ for $i \geq 1$.
Both directions of the proposition are proved separately for an initial state $\iget[state]{\I} \in \reals^n$.

``$\Rightarrow$''. Suppose $\iaccessible[\fastswitch]{\I}{\It}$.
Like the proof of~\rref{prop:adequacyslowswitching}, by \dL semantics, there are compacted sequences $(\tilde{\iget[state]{\I}}_0,\tilde{\iget[state]{\I}}_1,\dots,\tilde{\iget[state]{\I}}_n)$, $(\tilde{\zeta}_1,\dots,\tilde{\zeta}_n)$, and $(\tilde{p}_1,\dots,\tilde{p}_n)$ such that $ \tilde{\iget[state]{\I}}_0 = \iget[state]{\I}$, $\tilde{\iget[state]{\I}}_n = \iget[state]{\It}$, and $\tilde{\iget[state]{\I}}_{i-1}$ reaches $\tilde{\iget[state]{\I}}_{i}$ following the ODE $\D{x}=f_{\tilde{p}_i}{(x)}$ for $i \geq 1$.
Furthermore, $\tilde{p}_i \not= \tilde{p}_{i-1}$ for $i \geq 1$.
By definition of the controller $\alpha_u$ and domain constraints in $\fastswitch$, $\tilde{p}_i = \offmod{i}$ for $i \geq 1$, $\tilde{\zeta}_i = \zeta_{\offmod{i}}$ for $1 \leq i < n$, and $\tilde{\zeta}_n \leq \zeta_{\offmod{n}}$.
Consider the periodic switching signal $\sigma$ with switching times $\tau_i = \sum_{j=1}^{i} \zeta_{\offmod{j}}$ and the sequence of switching choices $p_i = \offmod{i}$ for $i \geq 1$.
By construction using~\rref{eqn:switching}, $\sigma$ is well-defined with the specified periodic switching times, and the solution $\solvar$ associated with $\sigma$ from $\iget[state]{\I}$ reaches $\iget[state]{\It}$ at time $\sum_{j=1}^{n} \tilde{\zeta}_j$.

``$\Leftarrow$''. Let $\sigma$ be the periodic switching signal with switching times $\tau_i = \sum_{j=1}^{i} \zeta_{\offmod{j}}$ and the sequence of switching choices $p_i = \offmod{i}$ for $i \geq 1$, and $\solvar : [0,\zeta) \to \reals^n$ be the associated switched system solution from $\iget[state]{\I}$.
Suppose the switched system reaches $\solvar(t)$ for $t \in [0,\zeta)$.
Let $\tau_{n}$ be the first switching time such that $t \leq \tau_{n}$; the index $n$ exists since $\tau_i \to \infty$.
Define the state sequence $\iget[state]{\I}_i \mnodefeq \solvar(\tau_i)$ for $0 \leq i < n$ and $\iget[state]{\I}_n = \solvar(t)$.
Note that $\iget[state]{\I}_0 = \iget[state]{\I}$ by definition of $\solvar(0)$.
By construction of $\solvar$, $\iget[state]{\I}_{i}$ is reached from $\iget[state]{\I}_{i-1}$ by following the solution to ODE $\D{x}=f_{p_i}(x)$ for exactly time $\zeta_{\offmod{i}}$ for $1 \leq i < n$ so switching is allowed by the controller $\alpha_u$ and domain constraints in $\fastswitch$.
$\hfill\qed$
\end{pf*}
\fi
\end{document}